\documentclass[10pt,twocolumn]{article} 
\usepackage{times,fullpage}
\usepackage{graphicx}
\usepackage{amssymb}
\usepackage{url,hyperref}
\usepackage{authblk}
\usepackage{todonotes}
\usepackage{tabularx}

\author[1]{Zheng Gong}
\author[2]{Carmine Ventre}
\author[1]{John O'Hara}
\affil[1]{Centre for Computational Finance and Economic Agents, University of Essex  \protect\\ Email: {\tt \{zg19500,johara\}@essex.ac.uk}}
\affil[2]{Department of Informatics, King's College London \protect\\ Email: {\tt carmine.ventre@kcl.ac.uk}}

\usepackage{multirow}
\usepackage[normalem]{ulem}
\useunder{\uline}{\ul}{}
\usepackage{subcaption}
\usepackage{algorithm}

\usepackage{amsmath} 
\DeclareMathOperator{\E}{\mathbb{E}}
\graphicspath{ {./Images/} }

\newcolumntype{Y}{>{\centering\arraybackslash}X}

\begin{document}
\title{The Efficient Hedging Frontier with Deep Neural Networks}

\date{}
\maketitle

\begin{abstract}
The trade off between risks and returns gives rise to multi-criteria optimisation problems that are well understood in finance, {efficient frontiers} being the tool to navigate their set of optimal solutions. Motivated by the recent advances in the use of deep neural networks in the context of hedging vanilla options when markets have frictions, we introduce the \emph{Efficient Hedging Frontier} (EHF) by enriching the pipeline with a filtering step that allows to trade off costs and risks. This way, a trader's risk preference is matched with an expected hedging cost on the frontier, and the corresponding hedging strategy can be computed with a deep neural network. 

We further develop our framework to improve the EHF and find better hedging strategies. By adding a random forest classifier to the pipeline to forecast market movements, we show how the frontier shifts towards lower costs and reduced risks, which indicates that the overall hedging performances have improved. In addition, by designing a new recurrent neural network, we also find strategies on the frontier where hedging costs are even lower. 
\end{abstract}


\section{Introduction} 
\label{Section: Intro}
In the past decades, the evolution of financial derivative markets has provided investors numerous opportunities for trading and, especially for managing risks associated with future commodity prices, stock prices, interest rates and exchange rates. The markets expanded massively in the past ten years \cite{Huan2019}. A vanilla option is one of the most basic financial derivatives. It gives the holder the right (not the obligation) to buy or sell an asset at a predetermined price \cite{Kwok2008}. In particular,  we will focus in this paper on European Call Options (as opposited to, e.g., American Option) wherein the buyer can only execute it at expiration. An investor utilises the option to benefit from a future price movement of the underlying asset which aligns with her expectation, and avoid the risk if the price moves in the opposite direction \cite{vittori2020}. The option buyer pays an option premium to the issuer at the inception of the contract, and both parties could frequently trade the underlying asset to hedge their exposures to the price movements. Therefore, finding a better solution for working out an appropriate option premium and generating hedging strategies are crucial. 

\begin{figure}
  \centering
  \subcaptionbox{Original Deep Hedging Pipeline\label{fig:DH}}{\includegraphics[width=0.45\textwidth]{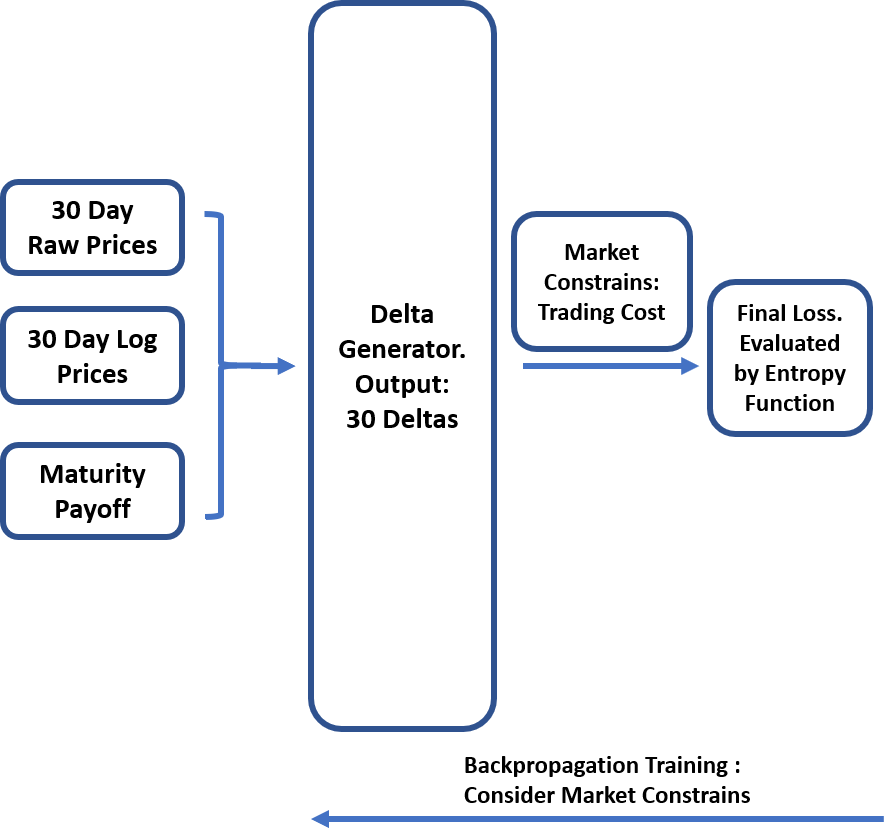}}\hfill%
  \subcaptionbox{Deep Hedging with Price Change Threshold\label{fig:DHWithPrice}}{\includegraphics[width=0.45\textwidth]{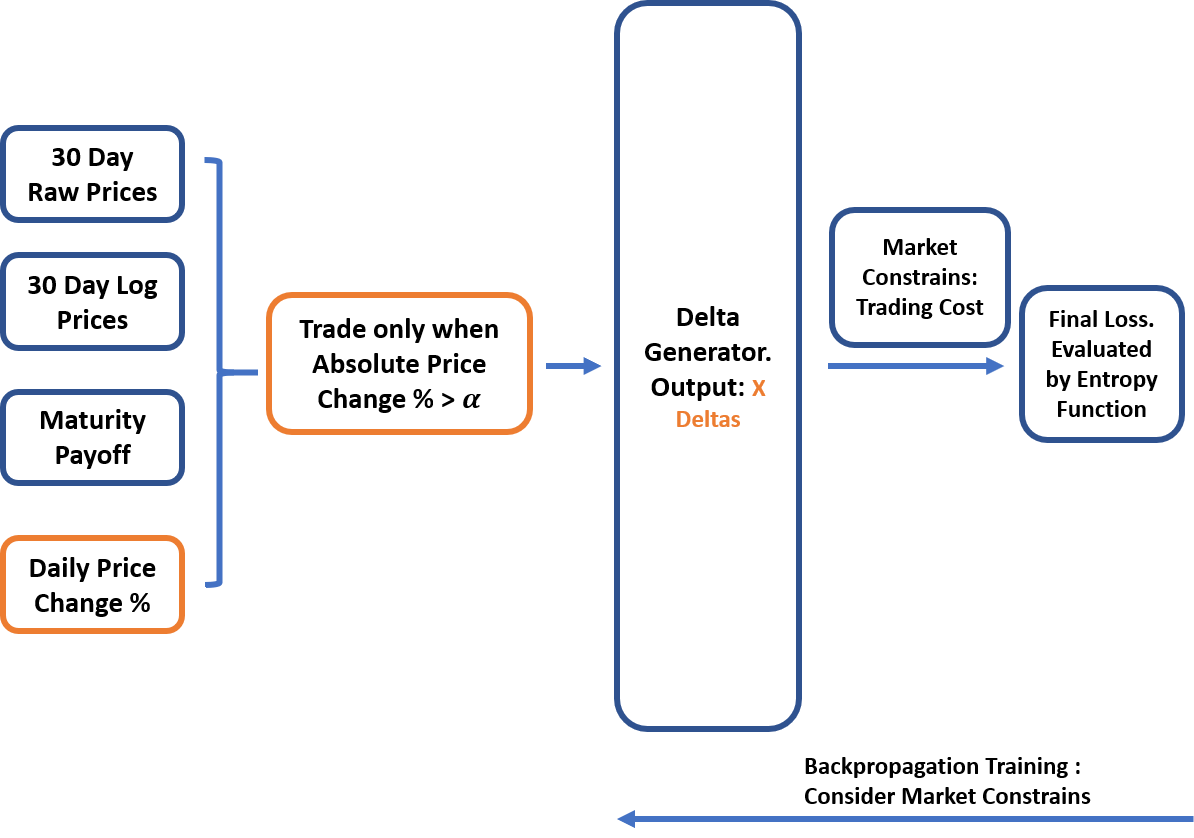}}\hfill%
  \subcaptionbox{Deep Hedging with Price Change Threshold and Random Forest\label{fig:DHWithRF}}{\includegraphics[width=0.45\textwidth]{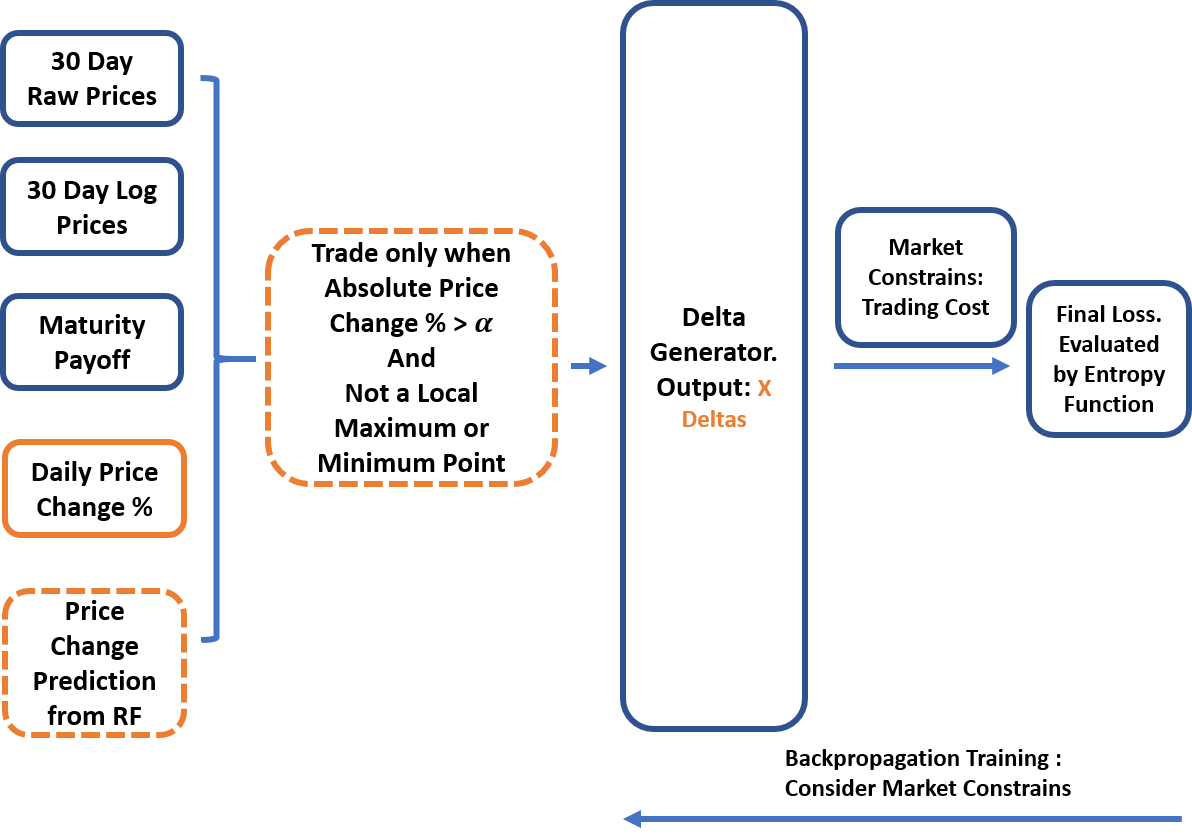}}%
    \caption{The Original and Amended Deep Hedging Pipelines}
\end{figure}

In early 1970s, Black and Scholes \cite{BlackScholes1973}, and independently, Merton \cite{Merton1973} initiated the classic parametric framework for option valuation and hedging, which we refer to as the Black-Scholes-Merton model. It is considered as a benchmark in every literature and widely applied in the industry. The model provides closed form solutions for pricing an option. Investors could also managing their hedging positions by calculating ``Greek Letters'', the partial derivatives of the value of an option with respect to the underlying asset price and other parameters in the Black-Scholes-Merton formula. Despite the popularity, the model is built on a set of idealised assumptions that are obviously not applicable in real life scenarios. First, the underlying price is modelled as a Geometric Brownian Motion (GBM) process with constant volatility. Therefore, it cannot model the fat tails of observed probability density, and gives rise to under estimated risks \cite{Boyarchenko2002}. Second, it assumes there are no trading costs and trading limitations for every participant in the market. Third, traders should continuously re-balance her positions in order to achieve zero profits (losses) at the maturity of the contract, which is financially and practically infeasible. Therefore, in reality, there is always a loss for the issuer at maturity of an option, so that the option premium could be calculated based on the expected losses. 

There are quite a lot articles discussing and proposing solutions to overcome the above-mentioned limitations of Black-Scholes-Merton framework. Heston \cite{Heston1993} and Bates process \cite{Bates1996} are two famous stochastic volatility models that introduce uncertainties in the behaviour of volatility, and consequently allow to simulate the price evolution of financial assets more realistically. Beginning with Hutchinson et al. \cite{Hutchinson1994}, neural network models were considered as a non-parametric solution to solve option pricing and hedging problems. It is believed that a neural network model has this typical capability of arbitrarily approximating any nonlinear relationship \cite{Hornik1991}. Deep Hedging \cite{Buehler2019} is one of the most recent advances in this line of work --- its technical pipeline is depicted in Figure \ref{fig:DH} for an option with 30 days maturity. The authors propose a framework to replicate conventional delta hedging strategies with learning networks. With Deep Hedging, traders could optimize models under different levels of transaction costs, as well as various risk measurements and risk appetites. It is also concluded that neural networks achieve better hedging performances than Black-Scholes-Merton model with real S\&P500 index data \textit{when re-calibrated on a daily basis}. There are also follow-ups stemming from this work \cite{vittori2020,Cao2021,Jang2021,Ruf2020}. These algorithms all incorporate neural network structures to solve the mapping between underlying price changes and optimal delta values, and their major differences are related to model architectures, loss functions and evaluation methods. 

However, there remains one inevitable question rarely asked: Why should a market participant always trade the underlying asset at a regular time interval (e.g. every day, every two days)? It is certainly not the case on the trading floor. Since continuous trading is impossible, traders often make decisions based on personal experience and knowledge. They decide the best timing to re-balance their hedging positions. For example, if it is believed that the underlying price is going to experience a V-shaped (or reverted V-shaped) pattern, then it is clearly a waste of money to sell some share and then, within a short period of time, buy it back as transaction costs would never be zero. 

In this work, we tackle the problem of letting the algorithm self-decide when it is the best moment to buy or sell the underlying asset. The decisions are based upon historical prices of the underlying, as well as the model's expectations about future prices movements. From the technical point of view, these two new inputs act as a \emph{filtering} step for the deep hedging network. 

We first introduce a price change threshold which restricts the model to perform trades only when the underlying prices experience significant movements, see Figure \ref{fig:DHWithPrice}. By changing the value of the price threshold $\alpha$, we get different incomparable hedging strategies. The larger $\alpha$ we use, the smaller the number of trades used by the strategy; consequently, we have smaller hedging costs (given the lower transaction fees) but bigger risk of experiencing a large loss at maturity (given that we hedge less effectively). More formally, costs and risks here are measured in terms of the mean and variance of the termination loss over a large number of market paths, respectively. We call the \textit{Efficient Hedging Frontier} (EHF) this curve of undominated strategies in the cost--risk space, inspired by the efficient frontiers defined in portfolio optimisation \cite{Markoitz1952} and algorithmic execution \cite{almgren2000optimal}. From the EHF, a market participant can pick a trade-off strategy which satisfies her risk and return preferences, by choosing an adequate value of $\alpha$. 

Our second filtering step is depicted in Figure \ref{fig:DHWithRF}. We additionally place a random forest classifier before the network to predict future movements of underlying prices. The classifier would instruct the neural network to hold its position if it believes a V-shaped pattern is coming. By adding the classifier, we could shift the EHF and obtain strategies where the mean and variance of termination losses are reduced simultaneously. 

Finally, we also experiment with the architecture of the hedging neural network and test the effectiveness of using recurrent architectures to leverage the temporal relationships in the price time series. We show how such a design choice can further shift the EHF towards even better strategies. 

The remainder of the paper is organised as follows. Section \ref{Section: Related Works} introduces related work on hedging with neural networks and efficient frontiers. Section \ref{Section: BSM} provides more details of the Black-Scholes-Merton framework. Section~\ref{Section: Heston} discusses the Heston stochastic volatility model as well as strengths and limitations of existing Deep Hedging models. Section \ref{Section: PriceChangeThreshold} describes our first setting which uses a predefined price change threshold to constrain the model trading activities. Section \ref{Section: DHwithClassifier} shows how a classifier could benefit a Deep Hedging neural network to reduce both hedging costs and risks. Section \ref{Section: Experiments} presents the results from our experiments and compares them with Black-Scholes-Merton and Deep Hedging. In section \ref{Section: Further Experiments}, we show our further experiments which replace dense networks with recurrent neural networks for Deep Hedging. Section \ref{Section: Conclusion} concludes our work and proposes directions for future research.

\section{Related Work} \label{Section: Related Works}
The first inspiration of pricing derivative asset with non parametric method started from Hutchinson et al. \cite{Hutchinson1994} in 1994. They compared simple multi-layer perception networks with other three popular methods (ordinary least squares, radial basis function networks and projection pursuit) for recovering the Black-Scholes-Merton formula of option prices with a time horizon of two years. Their work proved that learning networks could successfully generate option prices and delta-hedging strategies. More recently, as neural network models achieved predominant results in computer vision and natural language processing, these sophisticated models were also applied for option hedging problems. Cao et at. \cite{Cao2021} utilizes reinforcement learning framework to generate optimal hedging for a short position in a call option. They tried two different Q-functions to optimize mean of cost and variance of cost at the same time. They also compared two different evaluation methods (profit and loss approach, cash flow approach) to assess the hedging performances of their models. There is another reinforcement learning solution proposed in \cite{vittori2020}. It uses Trust Region Policy Optimization method to search the policy space and minimize the variance of rewards between one step and the next, in contrast to minimising one reward at the end of all steps. Ruf et al. \cite{Ruf2020} also created a neural network model to optimise average hedging error at the expiration of a call option, which is similar to profit and loss formulation in \cite{Cao2021}. 

Deep Hedging \cite{Buehler2019} (DH) is one of the most popular frameworks in this area. It not only considers market transaction costs when generating hedging solutions with neural networks, but more importantly, convex risk measures such as entropy risk function and expected shortfall can be applied with Deep Hedging. This is more relevant to industry practices than any other solutions mentioned before. As discussed in section \ref{Section: Intro}, there is a common shortcoming among all the existing methods which assumes trading at fixed intervals. We tackle the problem by filtering the trading days. 

The theory of optimizing investment strategies by considering the trade off between risk (volatility of profits) and return (the expected value of profits) came from Markowitz \cite{Markoitz1952}. If two portfolios of risky assets have the same return (or same risk), a rational investor would always prefer the one with lower risk (or higher return). Therefore on the risk--return plane, it is possible to draw an efficient frontier indicating all the optimal portfolios with different risk appetites. All the points under the frontier are not optimal, and if the frontier shifts up-left, the overall strategy gets better as both risk and return are improved. This intuitive concept is widely applied in many practical areas, such as for determining capital asset prices \cite{Perold2004}, algorithmic execution \cite{almgren2000optimal}, managing forward contracts \cite{Woo2004} and portfolio optimization \cite{Pla2013}. We adopt this idea of efficient frontier and apply it to Deep Hedging framework. By filtering trading days with large daily price changes, the algorithm re-balances its position at different trading frequencies. Larger price change thresholds lead to lower frequency, which means reduced trading costs but higher uncertainly of final cash value at option maturity. 

\section{Black-Scholes-Merton}
\label{Section: BSM}
First of all, a model is always required to reflect our assumptions for the price variations of the underlying asset. Geometric Brownian Motion (GBM) is a simple choice. Let \(S_t\) represents its price at time \(t\). We have:
\begin{equation} 
\label{eq: GBM}
dS_t = \mu S_t dt + \sigma S_t dB_t,
\end{equation}
where \(\mu\) is the drift, \(B_t\) is a Brownian motion which contributes uncertainty and \(\sigma\) controls volatility.

We also have an European call option contract with value of \(C_t(S_t,t)\). By using Itô's lemma \cite{Ito1944} on \(C_t\) and substituting (\ref{eq: GBM}), we obtain: 
\begin{equation} 
\label{eq: dCt}
\begin{split}
	dC_t  = & \left(\mu S_t \frac{\partial C_t}{\partial S_t} + \frac{\partial C_t}{\partial t} + \frac{1}{2} \sigma^2 S_t^2 \frac{\partial ^2 C_t}{\partial S_t ^2}\right) dt \\ & + \sigma S_t \frac{\partial C_t}{\partial S_t} dB_t.
\end{split}
\end{equation}

We want to replicate the option with a portfolio \(P\), which holds \(y_t\) of the underlying asset and \(x_t\) of risk-free assets. With risk-free rate \(r\), we have: 
\begin{align} 
 P_t &= x_t e^{rt} + y_t S_t  \label{eq: Pt} \\
 dP_t & = (r x_t e^{rt} + y_t \mu S_t)dt + \sigma y_t S_t dB_t. \label{eq: dPt}
\end{align}
We want any gains or losses on the portfolio \(P\) entirely due to the price movements of the underlying asset \(S_t\), and we can equate terms in (\ref{eq: dCt}) and (\ref{eq: dPt}) to obtain:
\begin{align} 
y_t & = \frac{\partial C_t}{\partial S_t} \label{eq: Delta} \\
\label{eq: xt}
x_t  & = \frac{\frac{\partial C_t}{\partial t} + \frac{1}{2} \sigma^2 S_t^2 \frac{\partial ^2 C_t}{\partial S_t ^2}}{re^{rt}}.
\end{align}
In an ideal world with no dividend, zero trading cost and unlimited capital, a trader could continuously adjust her portfolio \(P\) according to equation (\ref{eq: Delta}) and (\ref{eq: xt}) to always achieve zero profit or losses at the maturity of option. Equation (\ref{eq: Delta}) is effectively calculating the partial derivative of \(C_t\) with respect to \(S_t\), so this is usually called delta hedging strategy. 

We could further substitute equations (\ref{eq: Delta}) and (\ref{eq: xt}) back into (\ref{eq: Pt}) to obtain the famous Black-Scholes-Merton partial differential equation. 
\begin{equation}
\frac{\partial C_t}{\partial t} + rS_t\frac{\partial C_t}{\partial S_t} + \frac{1}{2}\sigma^2S_t^2\frac{\partial ^2 C_t}{\partial S_t ^2} - rC_t = 0.
\end{equation}
This equation can be solved by providing boundary conditions on \(C_t\), and therefore the option price at inception \(C_0\) together with hedging strategies \(x_t\) and \(y_t\) are established.

\section{Heston and Deep Hedging Framework}
\label{Section: Heston}
GBM process and delta hedging strategy work perfectly together under their idealized settings. To take one step closer to reality, we could use stochastic volatility models to simulate the underlying prices --- Heston model \cite{Heston1993} is selected in Deep Hedging \cite{Buehler2019} towards this end. We have: 
\begin{equation}
\begin{split}
dS_t &= \mu S_t dt + \sqrt{v_t} S_t dB_t^1 \\ dv_t &= \kappa (\theta - v_t) dt + \sigma \sqrt{v_t} dB_t^2.
\end{split}
\end{equation}

In the above stochastic differential equations, \(B_t^1\) and \(B_t^2\) are two one-dimensional Brownian motions, with correlation 
in \([-1,1]\); \(v_t\) controls the volatility of \(S_t\), which is a mean-reverting stochastic process instead of a constant. The parameter \(\sigma\) is called the volatility of the volatility which could be treated to model the general market environment, i.e., higher value represents more volatile markets. 

We use two sets of Heston parameters for our experiments, which try to simulate market scenarios with different levels of fluctuation. The values for those parameters are show in Table \ref{tb:Heston}.

\begin{table}[]
\centering
\caption{Heston parameters used in our experiments}
\label{tb:Heston}
\begin{tabular}{|c|c|c|c|c|c|c|}
\hline
Market Scenario & \(v_0\)  & \(\theta\) & \(\kappa\) & \(\mu\)   & \(\sigma\) & \(\rho\)  \\ \hline
Low Volatility  & 0.4 & 0.4   & 1     & 0.01 & 4     & -0.7 \\ \hline
High Volatility & 0.8 & 0.8   & 1     & 0.01 & 4     & -0.7 \\ \hline
\end{tabular}
\end{table}

The problem is to find the values of \(y_t\) in (\ref{eq: Pt}), so that \(P_t\) could replicate a call option on \(S_t\) as close as possible during the term of the contract. Delta hedging in (\ref{eq: Delta}) is no longer the perfect solution, because returns of the underlying prices \(S_t\) are not log-normally distributed with Heston. Deep Hedging generates the values of \(y_t\) (delta) using a neural network with two fully connected hidden layers. 
The core advantage of Deep Hedging is the \(y_t\) are obtained from training instead of pure mathematical calculations. They are the outputs from the two-layer delta generator which is trained end-to-end with an Adam optimizer \cite{Kingma2014}. In this way, not only practical limitations (e.g., trading costs) could be accounted for into the model, but more importantly, the model could be tailored with different risk functions and adjustable risk preferences. There are at least two risk functions proposed with Deep Hedging. One is the entropy risk measure:
\begin{equation}
\rho (X) = \frac{1}{\lambda}\log \E(e^{-\lambda X}).
\end{equation}
There is only one parameter \(\lambda\), which controls risk preferences; smaller \(\lambda\) indicates more risk aversion. During our experiments, we use this entropy risk measure as executed in \cite{Buehler2019}. 

\begin{table}[htb]
	\centering
	\caption{An illustration of Deep Hedging}
	\label{tb: DH}
	\begin{tabularx}{\columnwidth}{|c|c|c|c|Y|}
		\hline
		Day                 & Price              & DH Delta                 & Buy/Sell                 & Trading Cost                     \\ \hline
		0                   & 100.00                   & 0.4090                   & 40.9042                  & 2.0452                   \\ \hline
		\textbf{1}          & \textbf{100.13}          & \textbf{0.4092}          & \textbf{0.0178}          & \textbf{0.0009}          \\ \hline
		2                   & 106.12                   & 0.4334                   & 2.5711                   & 0.1286                   \\ \hline
		\textbf{3}          & \textbf{106.34}          & \textbf{0.4377}          & \textbf{0.4471}          & \textbf{0.0224}          \\ \hline
		\textit{\textbf{4}} & \textit{\textbf{109.43}} & \textit{\textbf{0.4559}} & \textit{\textbf{1.9992}} & \textit{\textbf{0.1000}} \\ \hline
		5                   & 106.71                   & 0.4704                   & 1.5435                   & 0.0772                   \\ \hline
		6                   & 102.52                   & 0.4711                   & 0.0684                   & 0.0034                   \\ \hline
		\textbf{7}          & \textbf{102.28}          & \textbf{0.5039}          & \textbf{3.3557}          & \textbf{0.1678}          \\ \hline
		8                   & 101.99                   & 0.4921                   & -1.2001                  & 0.0600                   \\ \hline
		\textit{\textbf{9}} & \textit{\textbf{105.46}} & \textit{\textbf{0.5205}} & \textit{\textbf{2.9990}} & \textit{\textbf{0.1500}} \\ \hline
		10                  & 103.59                   & 0.5114                   & -0.9491                  & 0.0475                   \\ \hline
	\end{tabularx}
\end{table}

The limitations of Deep Hedging are also quite obvious. The delta generator takes price information from every trading day, and outputs one best value of delta for that particular day. Sometimes, from one day to the next, the price change is negligible and the model buys or sells little amount of underlying asset. More importantly, when there is a V-shaped movement of underlying in two consecutive days, the model would sell some underlying and then buy them back, which causes unnecessary trading costs. This is illustrated in Table \ref{tb: DH}. At day 1, 3 and 7, the price changes are small comparing with the previous days. At day 4 and day 9, they are peak values of underlying prices. It is reasonable for a trader take no actions in these days; in this example, this will save roughly 15\% of trading costs in a 10-day period.

\section{Deep Hedging with a Price Change Threshold}
\label{Section: PriceChangeThreshold}

We first try to limit Deep Hedging to generate deltas only when the underlying price changes significantly from one day to another. Therefore, we introduce an additional input feature to the delta generator. As from above, this amended Deep Hedging pipeline is shown in figure \ref{fig:DHWithPrice}, where orange denotes our novel filtering.

The absolute percentage changes of daily prices are calculated from simulated trajectories, and only the days with absolute price changes higher than a predefined threshold \(\alpha\) will be considered by the neural network. For the other days, the deltas will remain unchanged, and therefore no buy or sell actions are taken. For the standard Deep Hedging algorithm, the model always outputs 30 deltas for each input path. By adding the threshold \(\alpha\), the trading frequency for each path reduces from 30 (daily trading) to 0 (no trading) as \(\alpha\) increase from 0 to roughly 0.3. We simulated 120,000 paths for our experiments, and Table (\ref{tb:alpha}) shows the average number of trading days for one trajectory when the value of \(\alpha\) varies. For example, if \(\alpha\) is set to 0.04, there would be only 9.53 trades performed during the 30-day period. 

\begin{table}[]
\centering
\caption{Trading frequency reduction as \(\alpha\) increases}
\label{tb:alpha}
\begin{tabular}{|c|c|}
\hline
Threshold \(\alpha\)  & Average Frequency \\ \hline
0.00      & 30.00         \\ \hline
0.02      & 16.64         \\ \hline
0.04      & 9.53          \\ \hline
0.06      & 5.20          \\ \hline
0.08      & 2.73          \\ \hline
0.10      & 1.40          \\ \hline
0.12      & 0.71          \\ \hline
0.14      & 0.36          \\ \hline
0.20      & 0.05          \\ \hline
\end{tabular}
\end{table}

\section{Deep Hedging with a Classifier}
\label{Section: DHwithClassifier}
A classifier is a model used to divide non-labelled data into different categories. It is very commonly applied with financial time series to predict future movements of asset prices. A decision tree is one of the most fundamental supervised classification model, which can be used to discover features and extract patterns for discrimination and predictive modelling \cite{Myles2004}. The idea is basically to break up a complex task into many simpler decisions, and for each decision, the algorithm tries to increase the homogeneity of each classification category. Random forest (RF) is a popular ensemble model of many decision trees, where each tree is trained with a sub-sample of the training dataset. The output is generated from votes of the trees and therefore could improve the predictive accuracy and control over-fitting \cite{breiman2001}. Our random forest has fifty trees and utilize Gini Impurity of decision measurement. 

Before running the Deep Hedging network, we first label our simulated daily prices with two labels. If one day's underlying price is higher (lower) than yesterday and lower (higher) than tomorrow by some threshold \(\beta\), we label it as zero, otherwise we label it as one. Basically, zero means do not trade one that day, because today's profit (loss) will be recovered tomorrow. We set \(\beta\) to be 0.05 for our experiments. We then take log-normalised prices from the previous two days and train a random forest classifier to classify every daily price into category zero or category one. Because we use synthetic data, the classification accuracy is relatively high with roughly 95\% for test data and 99\% for training data. Subsequently, we start training the Deep Hedging network, and add the labels predicted from the random forest classifier as an extra feature. These labels will instruct our neural network to skip the days, where underlying prices are local maxima or local minima, see Figure \ref{fig:DHWithRF}.

\section{Experimental Setting and Results}
\label{Section: Experiments}
Our neural network models are implemented in Python with Tensorflow. The random forest classifier utilised is from scikit-learn package \cite{scikit-learn}. We simulated 120,000 Heston trajectories for our experiments, split in 100,000 for training and 20,000 for testing. We train the network to optimise the issuer's accumulated cost at the maturity of an option contract, which we refer to as termination loss. In this section, we first present our experiments under high market volatility scenario, and then discuss the model performances under different market conditions. 

\begin{figure*}
\centering
\includegraphics[width=0.75\textwidth]{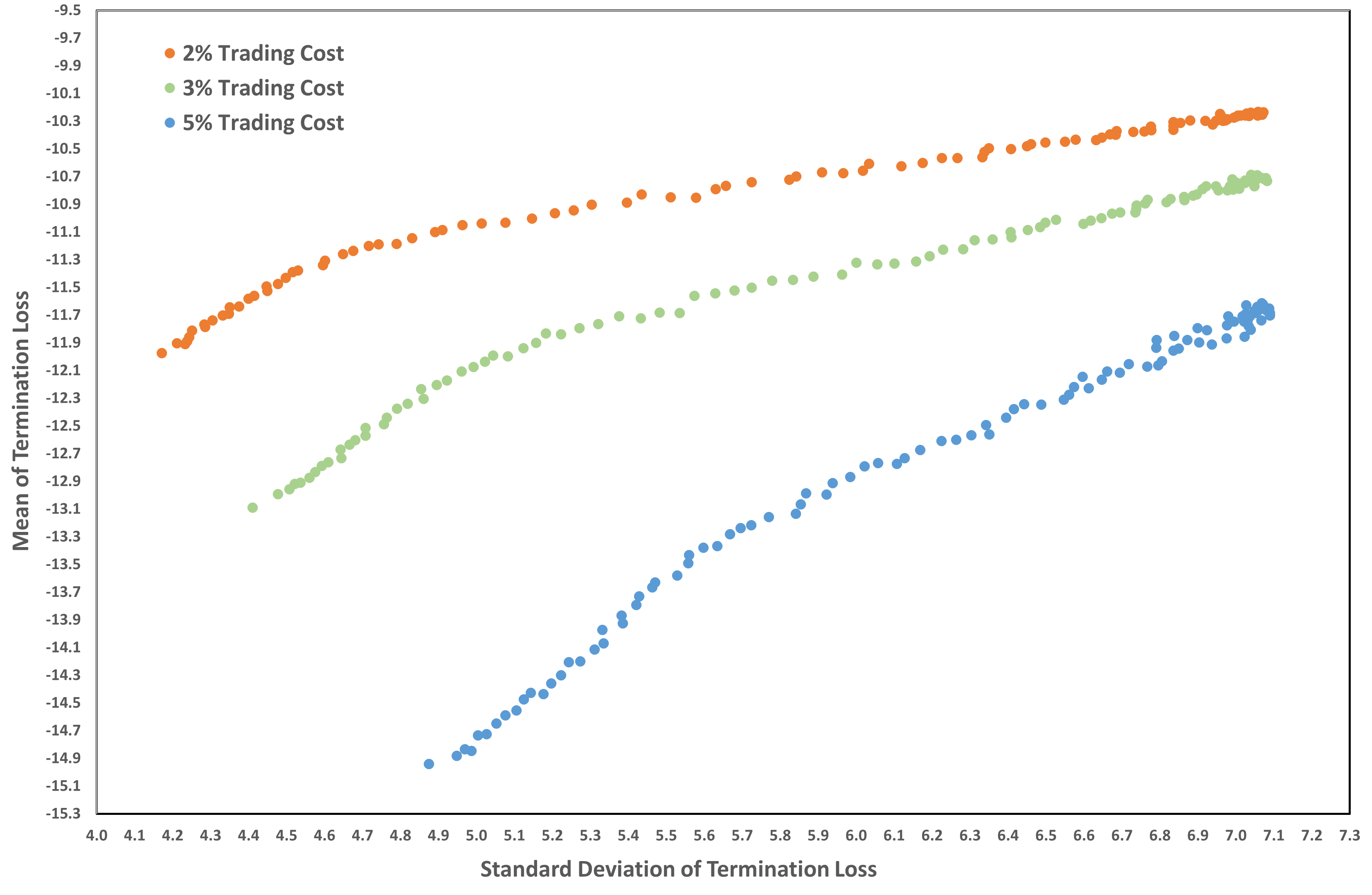}
\caption{The EHFs for different trading costs (\(\lambda = 0.5 \))}
\label{fig:Fron}
\end{figure*}

\begin{figure*}
	\centering
	\includegraphics[width=0.75\textwidth]{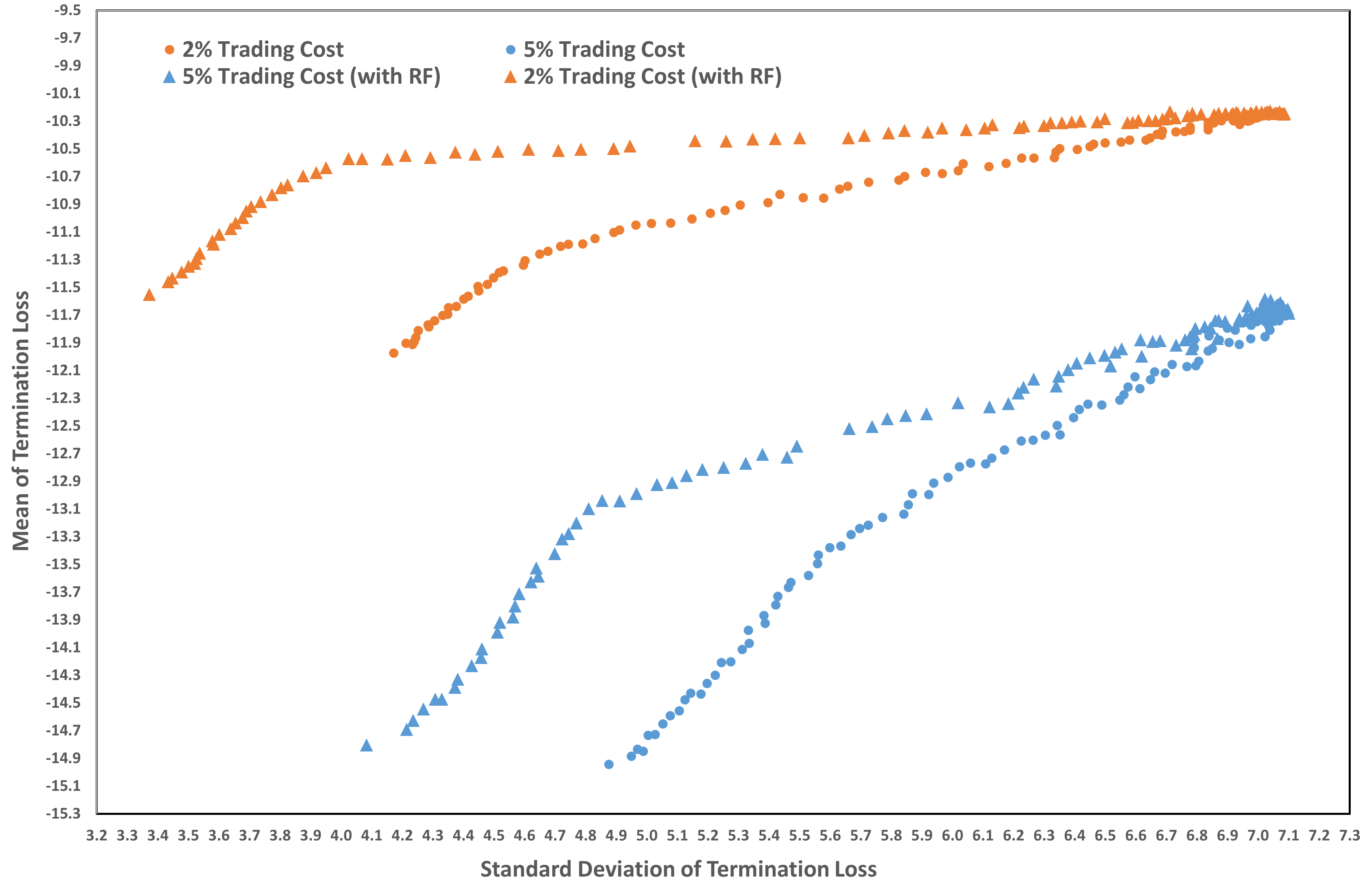}
	\caption{The EHFs with Random Forest forecast (\(\lambda = 0.5 \))}
	\label{fig:FronRF}
\end{figure*}

As mentioned in Section \ref{Section: Intro}, our ultimate objective is to reduce unnecessary trading for our Deep Hedging system. Using the approach discussed in Section \ref{Section: PriceChangeThreshold} we first attempt to force the network to only focus on trading days where there is a significant price changes. Comparing with the standard Deep Hedging \cite{Buehler2019}, there is one additional input feature of the daily price change percentage. The price change threshold \(\alpha\) could reduce the average termination losses for our 120,000 simulated paths, but also increase the standard deviation. Therefore, by tuning the value of \(\alpha\) we could obtain the EHF under high volatility market assumption as shown in Figure \ref{fig:Fron}.

There are three colours in Figure \ref{fig:Fron} indicating different market trading cost assumptions. Market costs are assumed to be proportional to the cash amount spent for buying/selling the underlying assets. There are 100 points for each line, and each point represents one particular price change threshold \(\alpha\) selected evenly from 0 to 0.2. The Y-coordinate of a point in Figure \ref{fig:Fron} is the mean of 20,000 termination losses from testing trajectories for a given value of \(\alpha\). The X-coordinate is the relevant standard deviation of these losses. As \(\alpha\) increases, the points move from left to right. Therefore, the bottom-left point is the standard Deep Hedging that trades every single day (\(\alpha = 0\)). At the top-right point of each line, where \(\alpha = 0.2\), the system is making only 0.05 trades during the 30-day period (see Table \ref{tb:alpha}). The average loss over 20,000 test paths gets very small (i.e. no trading cost), but the standard deviation of losses is significant (i.e. no hedging). It is also worthwhile to observe that at right end of each line, there are clusters of points. The explanation is that when \(\alpha\) makes small changes at high values (e.g., from 0.196 to 0.198), the algorithm could not filter out many extra trading days, so the results are faltering because of the randomness nature of neural networks. 

\begin{figure}[!b]
	\centering
	\includegraphics[width=0.45\textwidth]{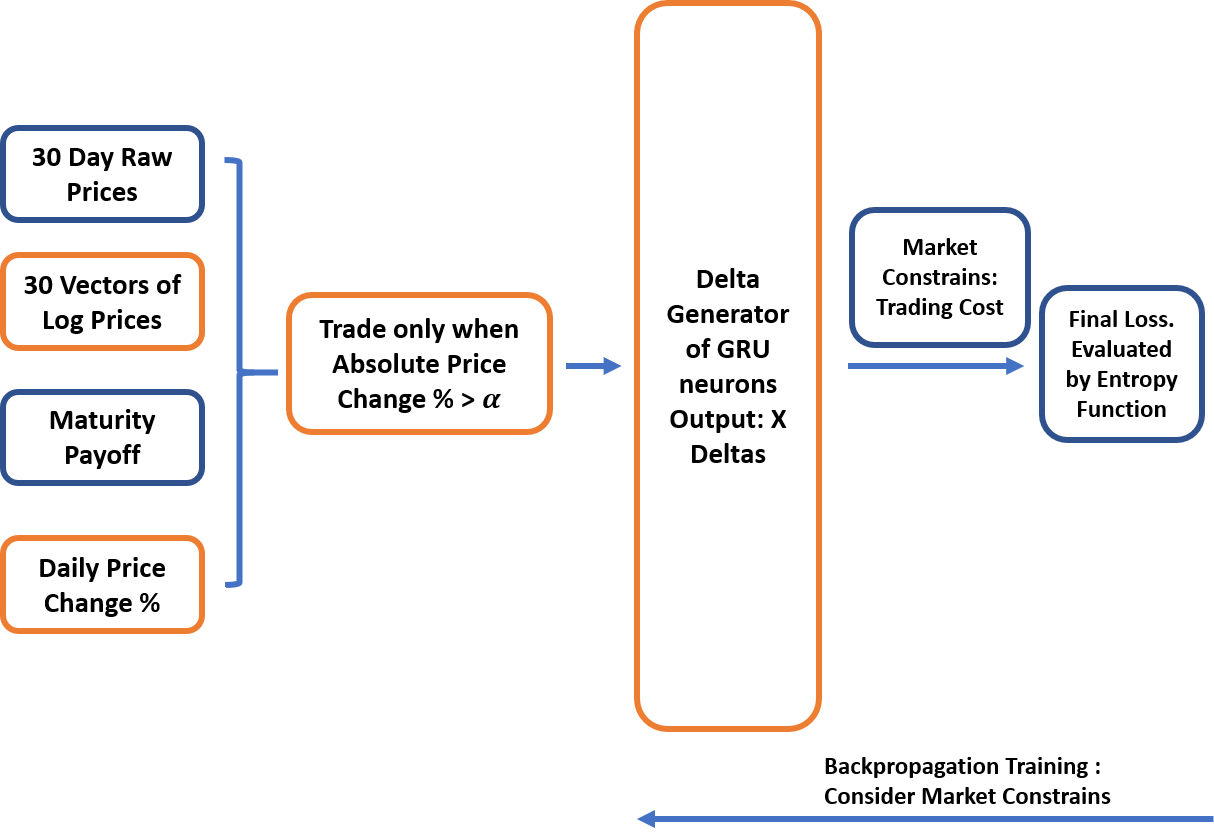}
	\caption{Deep Hedging using Gated Recurrent Network}
	\label{fig:DHGRU}
\end{figure}

At 5\% trading cost and \(\alpha \in [0, 0.1]\), we can calculate the average of mean termination losses as well as the average of standard deviations of termination losses from those points in in Figure \ref{fig:Fron}. The statistics are -13.628 and 5.578 respectively. If the hedging strategy is calculated with the Black-Scholes-Merton method instead, and average of means and average of standard deviations are -14.790 and 6.978 with the same values of \(\alpha\). This also proves Deep Hedging outperforms delta hedging by a clear margin with Heston simulations.

Clearly, adding a price change threshold is not actually improving Deep Hedging but provides a new prospective for trading-off risks and returns. An investor could decide a point on the efficient hedging frontiers to represent her risk appetite and then make the appropriate hedging strategy decisions. 

Our next step is to combine the Deep Hedging algorithm with a random forest classifier, as shown in Figure \ref{fig:DHWithRF}. The classification labels generated from the random forest is treated equivalently to the trader's expectations of the future movements of underlying prices. We show that if the classification task is solved sufficiently well (or, equivalently, the trader has good knowledge of the market) the hedging losses and risks could be reduced simultaneously. For high volatility market scenario, the result is shown in Figure \ref{fig:FronRF}. There are two groups of lines, which represent two trading cost assumptions. There are two lines in each colour group. The higher line shows the performances of Deep Hedging with the help of random forest classifier. At low values of \(\alpha\) (i.e. left end), the gap between performances of Deep Hedging with and without random forest is larger than at high values of \(\alpha\) (i.e. right end). When \(\alpha\) is really large, the two models exhibit similar performances, as \(\alpha\) is filtering out most trading days and good predictions could not make much contributions. 

\begin{table}[!t]
	\centering
	\caption{Improved Deep Hedging with Random Forest}
	\label{tb: RF}
	\begin{tabularx}{\columnwidth}{|c|c|c|c|Y|}
		\hline
		Day       & Stock Price      & DH Delta       & Buy/Sell       & Trading Cost     \\ \hline
		0          & 100.00          & 0.4334          & 43.3373         & 0.8667          \\ \hline
		1          & 97.09           & 0.4346          & 0.1144          & 0.0023          \\ \hline
		2          & 93.72           & 0.4300          & -0.4301         & 0.0086          \\ \hline
		\textbf{3} & \textbf{101.45} & \textbf{0.4300} & \textbf{0.0000} & \textbf{0.0000} \\ \hline
		4          & 93.91           & 0.4331          & 0.2969          & 0.0059          \\ \hline
		5          & 80.61           & 0.3064          & -10.2177        & 0.2044          \\ \hline
		6          & 82.60           & 0.3274          & 1.7344          & 0.0347          \\ \hline
		7          & 89.02           & 0.3803          & 4.7137          & 0.0943          \\ \hline
		\textbf{8} & \textbf{96.33}  & \textbf{0.3803} & \textbf{0.0000} & \textbf{0.0000} \\ \hline
		9          & 84.12           & 0.3122          & -5.7299         & 0.1146          \\ \hline
		10         & 83.97           & 0.3129          & 0.0603          & 0.0012          \\ \hline
	\end{tabularx}
\end{table}

\begin{figure*}
	\centering
	\includegraphics[width=0.75\textwidth]{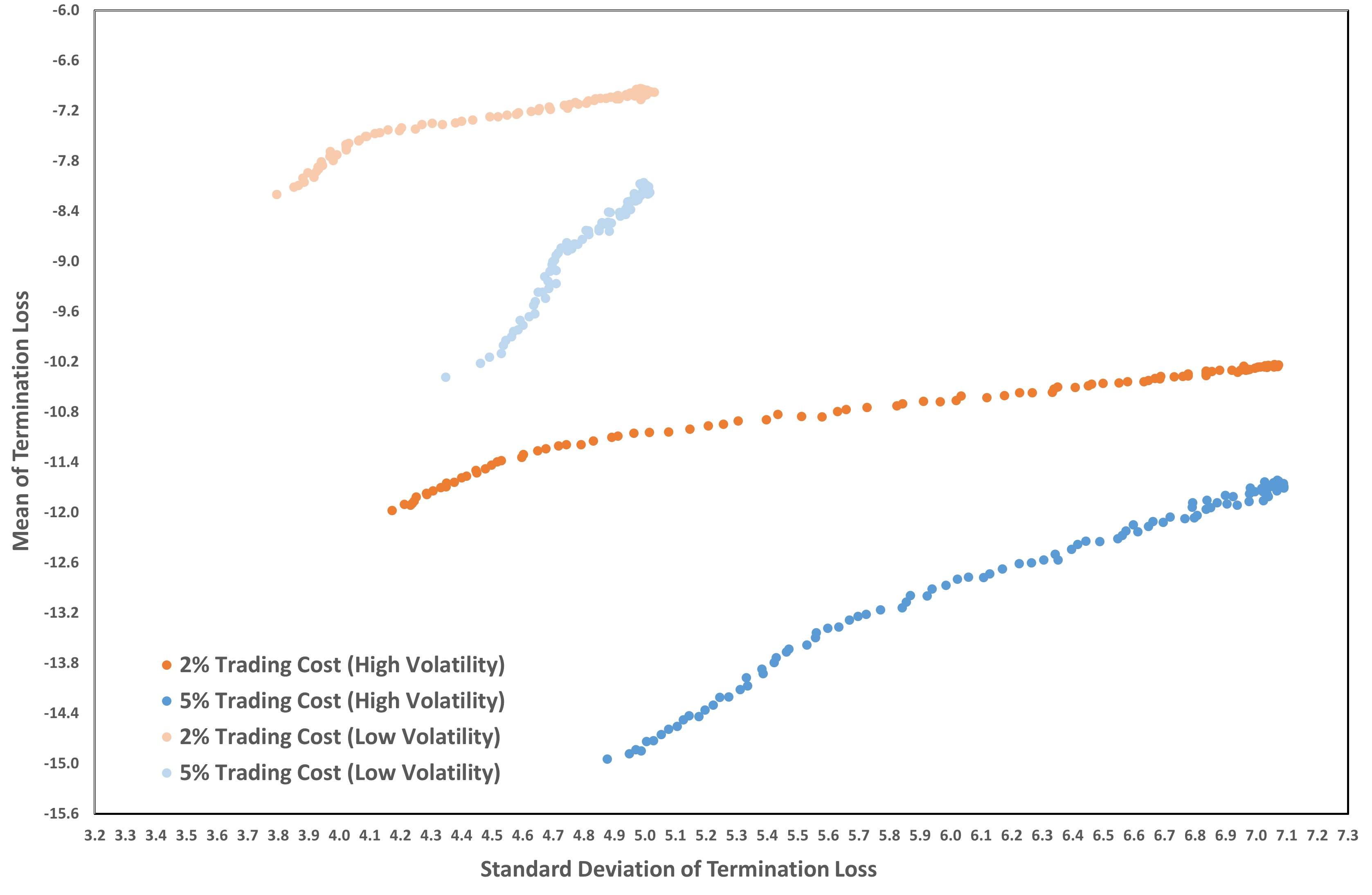}
	\caption{The Efficient Hedging Frontiers under different market conditions (\(\lambda = 0.5 \))}
	\label{fig:FronHighLow}
\end{figure*}

\begin{figure*}
	\centering
	\includegraphics[width=0.75\textwidth]{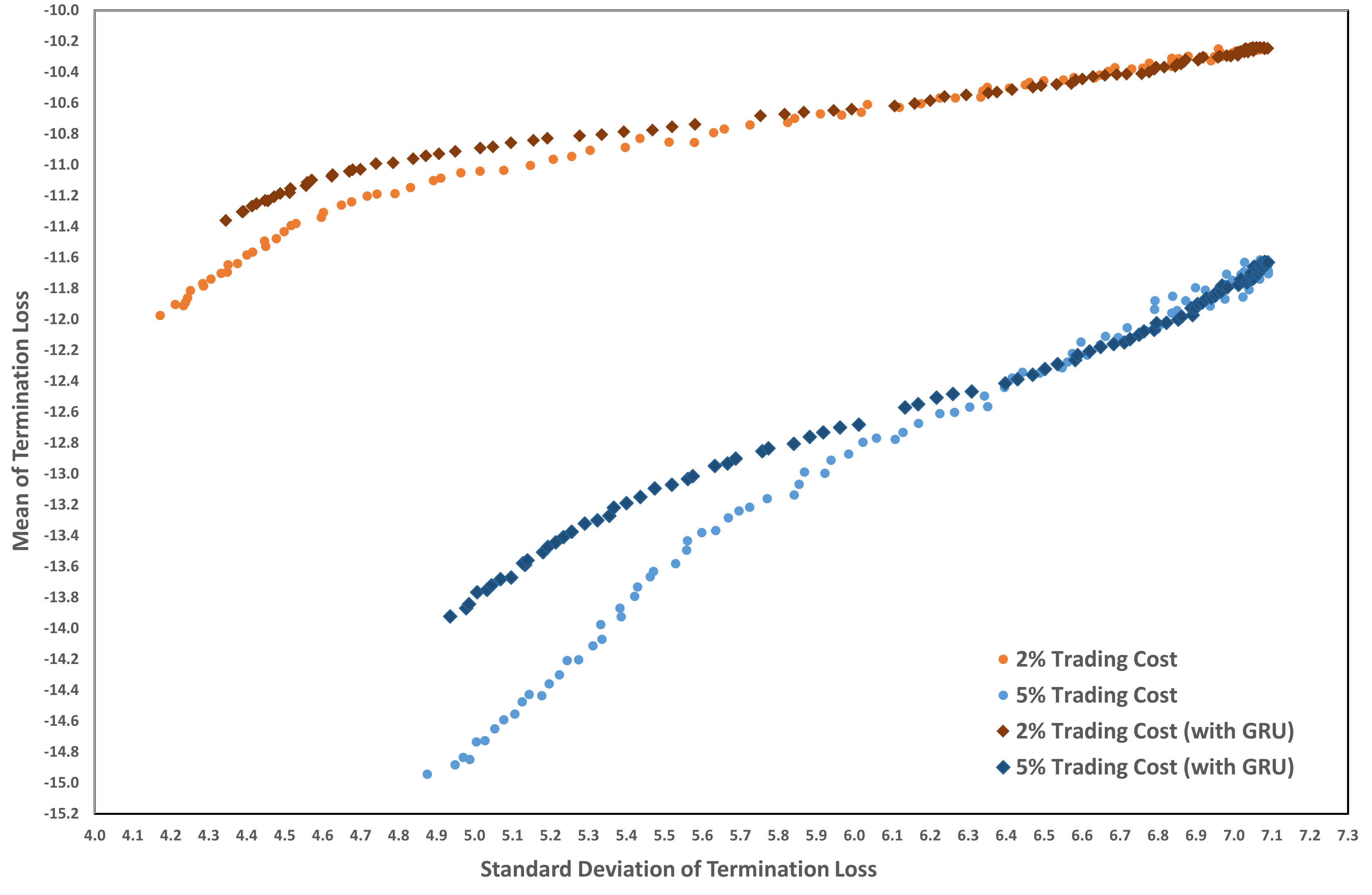}
	\caption{The Efficient Hedging Frontiers with GRU neural network (\(\lambda = 0.5 \))}
	\label{fig:FronGRU}
\end{figure*}

\begin{table*}[]
	\centering
	\caption{Improvement through RF classifier (\(\lambda = 0.5, \alpha \in [0, 0.1] \))}
	\label{tb:Compare0.5}
	\begin{tabular}{|c|c|c|c|c|c|c|}
		\cline{2-7}
		\multicolumn{1}{c|}{} & \multicolumn{3}{c}{Mean of Losses} \vline & \multicolumn{3}{c}{Standard Deviation of Losses} \vline \\  
		\cline{2-7} \multicolumn{1}{c|}{} & 2\% Cost & 3\% Cost & 5\% Cost & 2\% Cost & 3\% Cost & 5\% Cost  \\ \hline
		DH          & -11.253  & -11.898  & -13.628  & 4.887    & 5.143    & 5.578 \\ \hline
		DH with RF    & -10.718  & -10.784  & -11.683 & 4.209    & 4.353    & 4.543  \\ \hline
		Improvement & 4.75\%   & 9.36\%   & 14.27\%  & 13.87\%  & 15.37\%  & 18.54\% \\ \hline
	\end{tabular}
\end{table*}

Table \ref{tb: RF} gives an illustration of how the combined system makes hedging decisions. The forecasts from the random forest algorithm instructs the neural network that day 3 and day 8 are local maximum points for the underlying, according to its threshold (5\%), therefore the hedging generator skipped both days. The numerical comparisons of Deep Hedging with and without random forest classifier are shown in Table \ref{tb:Compare0.5}. 
Note we only take values of \(\alpha\) from 0 to 0.1 for calculating the averages, as larger values limit the trading frequencies too much. Overall, the improvement for standard deviations of termination losses is higher than for means of termination losses, which can also be visually observed in Figure \ref{fig:FronRF}.

We also test the model in the low volatility market condition; the EHFs are displayed in Figure \ref{fig:FronHighLow}. The price change thresholds \(\alpha\) considered are still in the interval from 0 to 0.2 and evenly distributed. It is noticed that when the underlying asset is less volatile, both mean and standard deviation of termination losses are reduced as expected. In addition, the length of the EHF is getting shorter and the points are more compactly distributed. It is also observed that the slope of the frontier is smaller with 2\% trading cost than with 5\% in both market conditions.

The above experiments are carried out with entropy risk measure parameter \(\lambda = 0.5\). If \(\lambda\) changes to 0.7, the frontier will shift slightly to the right and when it is 0.2, the frontier is slightly to the left. This is expected since the EHF moves in the same direction of the trader's risk aversion.

\begin{table*}[t]
	\centering
	\caption{Comparing neural network architectures (\(\lambda = 0.5, \alpha \in [0, 0.1]\))}
	\label{tb:CompareGRU}
	\begin{tabular}{|c|c|c|c|c|c|c|}
		\cline{2-7}
		\multicolumn{1}{c|}{} & \multicolumn{3}{c}{Mean of Losses} \vline & \multicolumn{3}{c}{Standard Deviation of Losses} \vline \\  
		\cline{2-7} \multicolumn{1}{c|}{} & 2\% Cost & 3\% Cost & 5\% Cost & 2\% Cost & 3\% Cost & 5\% Cost  \\ \hline
		DH           & -11.253  & -11.898  & -13.628 & 4.887    & 5.143    & 5.578 \\ \hline
		DH with GRU   & -10.938  & -10.685  & -13.094 & 5.073    & 5.259    & 5.597   \\ \hline
		Improvement  & 2.79\%   & 1.79\%   & 3.92\% & -3.18\%  & -2.27\%  & -0.35\% \\ \hline
	\end{tabular}
\end{table*}

\section{Updating the Neural Network}\label{Section: Further Experiments}
As discussed above, the default Deep Hedging model utilises two fully connected layers for the delta generator, and the input is only one daily price. Therefore, it does not consider the temporal relationships of underlying time series. It is very common and intuitive to choose recurrent neurons instead of dense connections for this problem. 

We tested the use of Gated Recurrent Unit (GRU) as the recurrent element and re-designed the Deep Hedging pipeline. As illustrated in Figure \ref{fig:DHGRU}, we use a vector (instead of a single number) to input historical prices in the past 3 days to the GRU layer. The delta generator consists of two recurrent layers each with ten recurrent units and one dense layer to output a single number, which means the optimal amount to hold the underlying asset. We need to point out that in the first two days for a trajectory, there are not enough past prices for constructing the vector, so for those we still incorporate dense layers as the default Deep Hedging. 

Comparing the EHFs of the standard Deep Hedging with
GRU version, we can conclude from Figure \ref{fig:FronGRU} that while the mean of termination losses are reduced with the GRU architecture for small values of \(\alpha\), the expected deviation of losses increases. The two lines overlap rather quickly as price change threshold \(\alpha\) gets larger. The numerical results are shown in Table \ref{tb:CompareGRU}. 

\section{Conclusions and Further Research} 
\label{Section: Conclusion}
We wanted to limit the trading activities of a Deep Hedging model so that unnecessary trading costs could be saved. By adding a price change threshold, which filters out trading days with insignificant price movements, we could generate an efficient hedging frontier. On the frontier, a market participant could intuitively balance her position between risk tolerances and expecting losses when hedging a European call option, and generate appropriate strategies accordingly. We experimented with various trading costs and market volatility assumptions, as well as different values of \(\lambda\) for entropy risk measures. We could also improve the efficient hedging frontier by incorporating a random forest classifier with the Deep Hedging neural network. Outputs from the classifier are treated as prior knowledge of how the underlying price will evolve in the near future, which helps the delta generator network to avoid trading against V-shaped movements. In addition, our experiments also proved that replacing dense layers with GRU layers could reduce the expected mean of termination losses for Deep Hedging, but increase the standard deviations. 

This research could be expanded to evaluate American options where the holder can exercise her right anytime before and including the contract expiration date. Heston model could be extended to Bates model to include volatility jumps for simulated trajectories. Real market data could also be explored to further test the robustness of efficient hedging frontier and the random forest classifier. The classification module could also be a learning network, so that delta generator and classifier could be trained end-to-end and simultaneously. 

\bibliographystyle{abbrv}

\bibliography{mybib}
\end{document}